\documentclass[authoryear,12pt]{article}

\usepackage{natbib}
\usepackage[title]{appendix}
\usepackage[english]{babel}
\usepackage{amsmath}
\usepackage{latexsym}
\usepackage{amssymb}
\usepackage{amsmath}
\usepackage{amsthm}
\usepackage{mathtools}
\usepackage{multirow}
\usepackage{multicol}
\usepackage{float}
\usepackage{graphicx}
\usepackage{booktabs}
\usepackage{hyperref}
\usepackage{array}
\usepackage{hyperref}
\usepackage{enumitem}
\usepackage{colortbl}
\usepackage{subfig}
\usepackage{epsfig}
\usepackage{lscape}
\usepackage[normalem]{ulem}
\usepackage[usenames,dvipsnames]{xcolor}
\usepackage{tikz}
\usetikzlibrary{bayesnet}
\usepackage{bbm}
\usepackage{caption} 
\usepackage{rotating}
\usepackage{algorithm}
\usepackage{algpseudocode}
\usepackage[blocks]{authblk} % for authors informations
\makeatletter
\newcommand*{\centernot}{%
	\mathpalette\@centernot
}
\def\@centernot#1#2{%
	\mathrel{%
		\rlap{%
			\settowidth\dimen@{$\m@th#1{#2}$}%
			\kern.5\dimen@
			\settowidth\dimen@{$\m@th#1=$}%
			\kern-.5\dimen@
			$\m@th#1\not$%
		}%
		{#2}%
	}%
}
\makeatother

\newcommand{\vertiii}[1]{{\left\vert\kern-0.20ex\left\vert\kern-0.20ex\left\vert #1 
	\right\vert\kern-0.20ex\right\vert\kern-0.20ex\right\vert}}

	\newtheorem{defin}{Definition}

\newcommand{\R}{\mathbb{R}}

\newcommand{\Fb}{\mathcal{F}}

\hypersetup{
colorlinks=true,
linkcolor=blue,
filecolor=blue,      
urlcolor= blue,
citecolor=blue,
}

\begin{document}

\title{Inference for Network Count Time Series	with the R Package PNAR}

\author[1,2]{Mirko Armillotta \footnote{Email address: m.armillotta@vu.nl.}}
\author[3]{Michail Tsagris}
\author[4]{Konstantinos Fokianos}

\affil[1]{Department of Econometrics and Data Science, Vrije Universiteit Amsterdam, The Netherlands} \affil[2]{Tinbergen Institute, Amsterdam, The Netherlands}
\affil[3]{Department of Economics, University of Crete, Greece}
\affil[4]{Department of Mathematics and Statistics, University of Cyprus, Cyprus}

{
	\makeatletter
	\renewcommand\AB@affilsepx{: \protect\Affilfont}
	\makeatother
	
	%\affil[ ]{Email}
	
	\makeatletter
	\renewcommand\AB@affilsepx{, \protect\Affilfont}
	\makeatother
	
	%	\affil[a]{m.armillotta@vu.nl}
	%\affil[b]{p.gorgi@vu.nl}
}

{\let\newpage\relax\maketitle}

\vspace{-0.1cm}

\begin{abstract}
		\noindent We introduce a new R package useful for inference about network count time series. Such data are frequently encountered in statistics and they are usually treated as multivariate time series. Their statistical analysis is based on linear or log-linear models.  Nonlinear models, which have been applied successfully in several research areas, have been neglected from such applications mainly because of their computational complexity. We provide R users the flexibility to fit and study nonlinear network count time series models which include either a drift in the intercept or a regime switching mechanism. We develop several computational tools including estimation of various count Network Autoregressive models and fast computational algorithms for testing linearity in standard cases and when non-identifiable parameters hamper the analysis. Finally, we introduce a copula Poisson algorithm for simulating multivariate network count time series. We illustrate the methodology by modeling weekly number of influenza cases in Germany.
\end{abstract}

%\doublespacing

\section{Introduction}
\label{sec:Introduction}

Many data examples   are frequently observed as  multivariate   counting processes recorded over a time-span and with known relations among  observations (e.g epidemiological data with geographical distances between different areas). 
An important  objective   then  is to study  the effect of a  known network to the observed data. 
This  motivates a great amount of  interest to  network  time series models; see 
\cite{zhu2017}  who developed continuous  Network Autoregressive models (abbreviated as NAR). 
For these models,  the observed  variable $Y$, for the node $i$ at time $t$, is denoted by $Y_{i,t}$, and it is assumed to depend   on the past value of the variable for the node itself,  say $Y_{i,t-1}$, 
and of the past values of the average of all its neighboring variables. The unknown parameters of the model are estimated by Least Squares estimation (LS). This work was advanced  by 
\cite{armillotta_fokianos_2021} who developed  linear and  
log-linear Poisson Network Autoregression model (PNAR) for multivariate count  distributed data. 
The joint dependence among different variables is specified by a copula construction \cite[Sec.~2]{fok2020}. In addition,  
\cite{armillotta_fokianos_2021} have further established 
parametric estimation under the framework of quasi maximum likelihood inference (see \cite{wedderburn1974quasi}, \cite{gourieroux1984pseudo}) and associated asymptotic theory when the network dimension increases. For the linear model case and in the context of epidemiology,  
related applied work has been developed  by \cite{held_etal_2005} and it was  extended by \cite{paul_etal_2008,paul_etal_2011,held_etal_2012,Meyer_etal_2014} and \cite{bracher_2020endemic}.

The previous contributions impose  linearity (or log-linearity) of the model,
which can be  a restrictive assumption for real world applications.  For example, 
existence of different underlying states  (e.g. exponentially expanding pandemic/ dying out pandemic)  implies that  different regime switching data generating processes should  be applied  and this fits the framework
we consider. Recently,  \cite{armillotta_fokianos_2022_testing} specified a general nonlinear Network Autoregressive model for both continuous and discrete-valued processes, establishing also related theory. In addition, the authors study  testing procedures for examining linearity of NAR model against specific nonlinear alternatives by means of a quasi score test statistic.
This methodology was developed  with and without the presence of  identifiable parameters under the null hypothesis.

Even though there exists sufficient statistical methodology for PNAR models,  there has been a lack of up-to-date software for implementing their analyses. The aim of this work is to fill this gap by introducing  the new package \texttt{PNAR} \citep{armillotta_et_al_rpackage_2022} and to 
demonstrate its usefulness  for count  network data analysis. 
Related R packages    do not provide  tools  for estimating  nonlinear models and  applying associated testing procedures. The package \texttt{GNAR} \citep{gnar_rpackage,Knightetal(2020)} studies Generalized NAR models (GNAR); this is a linear NAR model which takes into account  the effect of several layers of connections between the nodes of the network. This package deals  with  continuous-valued time series and does not contain 
tools   for testing linearity. \texttt{PNAR} complements \texttt{GNAR} as it provides additional
methodology for testing and inference about nonlinear  discrete-valued network models. 

Package \texttt{surveillance} \citep{sureveillance_rpackage, Meyer_etal_2017}  fits 
only linear  models for  spatial-temporal  disease counts with Poisson or Negative Binomial distribution and with an autoregressive network effect.  The package does accommodate various structural break-point  tests but it  does not contains functions for testing linearity and for log-linear model fitting. Moreover, standard errors of estimated parameters are computed by considering the quasi-likelihood as the true likelihood of the model \cite[Sec.~2.3]{paul_etal_2008}. However, if the count time series are cross-sectional dependent, as it is usually the case,  the likelihood function  is misspecified and the obtained  standard errors are not consistently estimated.

The \texttt{PNAR}  package provides several advancements to the state of the  art  software: i) efficient estimation of (log-)linear models with proper robust standard errors accounting for possible model mispecification;  ii) appropriate functions for testing linearity when a parameter is either identifiable of non-identifiable, under then null hypothesis, by providing appropriate  $p$-value bounds and bootstrap approximations; iii)  new algorithms for generating (log-)linear and nonlinear network count time series models.

The paper is organized as follows. The next \hyperref[sec:Network autoregressive models]{section}
introduces linear and log-linear network time series autoregressive models for count data. Details about inference for unknown 
model parameters  are  provided. An application to estimation of weekly number of influenza A \& B cases from two Southern German states is given and some further model  aspects are discussed.
Then we focus on non-linear models and associated testing theory. 
Results concerning score tests for testing linearity in NAR models are discussed and applied to influenza  data. We also address the issue of computational speed.   A short section discussing   simulation of  network count time series shows the usefulness of this methodology.  The paper concludes with a short discussion.

\section{Poisson network models}
\label{sec:Network autoregressive models}

Consider a known  network with $N$ nodes, indexed by $i=1,\dots N$.  
The neighborhood structure of such a network is completely described by its adjacency matrix, say $A=(a_{ij})\in\R x^{N\times N}$ where  $a_{ij}=1$, if there is a directed edge from $i$ to $j$, say $i\to j$, and 0 otherwise. 
Undirected graphs are allowed ($A=A^\prime$), which means that the edge between two nodes, $i$ and $j$, has no specific direction (say $i \sim j$). This is common in geographical and epidemic networks (e.g. district $i$ shares a border with district $j$, patient $i$ has a contact with patient $j$). 
Self-relationships are excluded i.e. $a_{ii}=0$ for any $i=1,\dots,N$. 

Let $Y$ be  a count variable measured on each node of the network ($i=1,\dots, N$), over a window of time ($t=1,\dots,T$). The
data is a $N$-dimensional vector of time series $Y_t=(Y_{1,t},\dots,Y_{i,t},\dots Y_{N,t})^\prime$, which is observed over the domain $t=1,2\dots,T$; in this way, a univariate time series is observed for each node, say $Y_{i,t}$, with corresponding conditional expectation $\lambda_{i,t}$. Denote by
$ \lambda_t=\mathrm{E}(Y_t|\Fb_{t-1})$ with $\lambda_t=(\lambda_{1,t},\dots,\lambda_{i,t},\dots,\lambda_{N,t})^\prime$ the conditional expectation vector of the counts with respect to their past history  $\Fb_{t-1}$.
The following linear autoregressive network model  takes  into account the known relations between nodes
\begin{equation}
	Y_{i,t}|\mathcal{F}_{t-1}\sim Poisson(\lambda_{i,t}), ~~~ \lambda_{i,t}=\beta_0+\beta_1n_i^{-1}\sum_{j=1}^{N}a_{ij}Y_{j,t-1}+\beta_2Y_{i,t-1}\,,
	\label{pnar_1}
\end{equation}
where $n_i=\sum_{j\neq i}a_{ij}$ is the total number of connections starting from the node $i$, such that $i\to j$; called out-degree. We call \eqref{pnar_1} linear Poisson Network Autoregression of order 1, abbreviated by PNAR(1); \citep{armillotta_fokianos_2021}. From the left hand side equation of \eqref{pnar_1}, we observe that the process $Y_{i,t}$ is assumed to be marginally Poisson but the joint process depends upon a copula function described in \hyperref[sec: Generate data]{simulations} at the end of the paper.  
Note that $\beta_0,\beta_1,\beta_2 >0$ since the conditional mean of the Poisson is positive.
Model \eqref{pnar_1} postulates that, for every single node $i$, the marginal  conditional mean of the process is regressed on:
\begin{itemize}
	\item the average count of the other nodes $j\neq i$ which have a connection with $i$; the parameter $\beta_1$ is called network effect, as it measures the average impact of node $i$'s connections;
	\item  the past count of the variable itself for $i$;  the coefficient $\beta_2$ is called autoregressive effect because it provides an estimator  for the impact of past count $Y_{i,t-1}$.
\end{itemize}
Model \eqref{pnar_1} implies  that only  nodes directly followed by the focal node $i$ (i.e. $i \to j$),  possibly, have an impact on its mean process of counts. It is a reasonable assumption in many applications; for example, in a social network the activity of node $k$, which satisfies  $a_{ik}=0$, does not affect node $i$. Hence, \eqref{pnar_1} measures the effect of a network to the observed multivariate count time series. 
Moreover, the model accommodates different types of network connectivity i.e. $a_{i,j}$ does not necessarily take the values  $1$-$0$ (connected-not connected). For example,  $a_{i,j}=1/d_{i,j}$ where $d_{i,j}$ is some measure of distance between node $i$ and node $j$ and  $a_{i,i}=0$. In this way the network effect becomes a spatial network component; see  the last  paragraph of \citet[p.3]{Knightetal(2020)} for a discussion about a similar set of weights.

More generally, the counts $Y_{i,t}$ can be assumed to depend on the last $p$ lagged values and $q$ covariates. Then consider the PNAR($p,q$) model
\begin{equation}
	\lambda_{i,t}=\beta_0+\sum_{h=1}^{p}\beta_{1h}\left( n_i^{-1}\sum_{j=1}^{N}a_{ij}Y_{j,t-h}\right) +\sum_{h=1}^{p}\beta_{2h}Y_{i,t-h} + \sum_{l=1}^{q}\delta_lZ_{i,l}\,,
	\label{pnar_pq}
\end{equation}
where $\beta_0, \beta_{1h}, \beta_{2h} \geq 0$, for all $h=1\dots,p$, $\delta_{l} \geq 0$, $l=1,2,\dots,q$  and $Z_{i,l}$ are non-negative covariates measured for each node $i=1,\dots,N$. If $p=1$ and $q=0$ set $\beta_{11}=\beta_1$, $\beta_{21}=\beta_2$ to obtain \eqref{pnar_1}. Model \eqref{pnar_pq} is stationary if $\sum_{h=1}^{p}(\beta_{1h}+\beta_{2h})<1$ \citep{armillotta_fokianos_2021}.

The linear form of \eqref{pnar_pq} offers  a great advantage interpreting  the parameters but  it accommodates  positive covariates.  A real valued covariate enters  \eqref{pnar_pq}  
through suitable transformations that  ensure positivity (e.g. include $\exp(Z)$ instead of directly $Z$). This restriction is bypassed by the log-linear model \citep{armillotta_fokianos_2021}:
\begin{equation}
	\nu_{i,t}=\beta_0+\sum_{h=1}^{p}\beta_{1h}\left(n_i^{-1}\sum_{j=1}^{N}a_{ij}\log(1+Y_{j,t-h})\right) +\sum_{h=1}^{p}\beta_{2h}\log(1+Y_{i,t-h}) + \sum_{l=1}^{q}\delta_lZ_{i,l} \,,
	\label{log_pnar_pq}
\end{equation}
where $\nu_{i,t}=\log(\lambda_{i,t})$ and the observation are still marginally Poisson, $	Y_{i,t}|\mathcal{F}_{t-1}\sim Poisson(\exp(\nu_{i,t}))$, for every $i=1,\dots,N$. Then the  model parameters are real-valued  since $\nu_{i,t}\in\R x$ and the covariates can take any real values. The stationarity condition turns out to be $\sum_{h=1}^{p}(|\beta_{1h}|+|\beta_{2h}|)<1$. Moreover, the interpretation of coefficients is similar to the case of linear model \eqref{pnar_1} but on the log-scale.

\subsection{Inference} \label{subsec:Inference}

Model \eqref{pnar_pq}, or \eqref{log_pnar_pq}, depends on the  $m$-dimensional vector of unknown parameters $\theta=(\beta_0, \beta_{11},\dots, \beta_{1p},\break \beta_{21}, \penalty 0 \dots, \beta_{2p}, \delta_1,\dots,\delta_q)^\prime$, with $ m=1+2p+q$. We use  of quasi-maximum likelihood methodology
for estimation of $\theta$;  see \cite{wedderburn1974quasi} and \cite{gourieroux1984pseudo}. The Quasi Maximum Likelihood Estimator (QMLE) is the vector of parameters $\hat{\theta}$ maximizing the function
\begin{equation}
	l_{T}(\theta)=\sum_{t=1}^{T}\sum_{i=1}^{N}\Big(  Y_{i,t}\log\lambda_{i,t}(\theta)-\lambda_{i,t}(\theta)\Big) \,,
	\label{pois-log-lik}
\end{equation}
which is the so called pooled Poisson log-likelihood (up to a constant). Note that \eqref{pois-log-lik}
is not necessarily the \emph{true} log-likelihood of the process but it serves as an approximation. In particular, 
\eqref{pois-log-lik} is the log-likelihood function that would have been obtained if all time series were contemporaneously independent.  However, the QMLE is not computed under the assumption of independence because  \eqref{pois-log-lik} is  simply a  \emph{working} log-likelihood function.  
The choice of maximizing \eqref{pois-log-lik} is justified for several reasons: 
i) full likelihood based on the joint process %as described in Section~\ref{subsec: Generate data}
is complex (see the last \hyperref[sec: Generate data]{section});  
ii) the optimization of \eqref{pois-log-lik} guarantees consistency and asymptotic normality of QMLE  for the \emph{true} parameter vector  $\theta_0$; iii)  the QMLE is asymptotically equivalent to the MLE if the true probability mass function belongs to the linear exponential family \citep{gourieroux1984pseudo}; 
iv) simplified computations entailing increased speed for estimation.
Robustness of the QMLE in finite samples has been verified by \cite{armillotta_fokianos_2021} through extensive simulation studies. 

When considering the linear model \eqref{pnar_pq}, the score function is 
\begin{equation}
	S_{T}(\theta)=\sum_{t=1}^{T}\sum_{i=1}^{N}\left(  \frac{Y_{i,t}}{\lambda_{i,t}(\theta)}-1\right) \frac{\partial\lambda_{i,t}(\theta)}{\partial\theta}
	=\sum_{t=1}^{T}s_{t}(\theta)
	\,.
	\label{score_poisson}
\end{equation}
Define $\partial\lambda_t(\theta)/\partial\theta^\prime$ the $N\times m$ matrix of derivatives, $D_t(\theta)$ the $N\times N$ diagonal matrix with elements equal to $\lambda_{i,t}(\theta)$, for $i=1,\dots,N$ and $\xi_t(\theta)=Y_t-\lambda_t(\theta)$ is the error sequence. Then, the empirical Hessian and conditional information matrices are given, respectively, by
\begin{align}
	H_{T}(\theta)=\sum_{t=1}^{T}\sum_{i=1}^{N}\frac{Y_{i,t}}{\lambda_{i,t}^2(\theta)}\frac{\partial\lambda_{i,t}(\theta)}{\partial\theta}\frac{\partial\lambda_{i,t}(\theta)}{\partial\theta^\prime}\,,
	~~~~
	B_{T}(\theta)=\sum_{t=1}^{T}\frac{\partial\lambda^\prime_t(\theta)}{\partial\theta}D^{-1}_t(\theta)\Sigma_t(\theta) D^{-1}_t(\theta)\frac{\partial\lambda_t(\theta)}{\partial\theta^\prime}\,,
	\label{H_T} 
\end{align}
where $\Sigma_t(\theta)=\mathrm{E}\left( \xi_t(\theta)\xi_t^\prime(\theta)\left| \right. \mathcal{F}_{t-1} \right) $ is the conditional covariance matrix evaluated at $\theta$. Under suitable assumptions, \cite{armillotta_fokianos_2021} proved  that  $ \sqrt{NT}(\hat{\theta}-\theta_0)\xrightarrow{d}N(0,H^{-1}BH^{-1})$, when $N\to\infty$ and $T\to\infty$, where $H$ and $B$ are the theoretical limiting Hessian and information matrices, respectively, evaluated at the true value  $\theta=\theta_0$.
Then, a suitable estimator for the standard errors of $\theta$ is the square-rooted main diagonal of the empirical "sandwich" covariance matrix, i.e. $SE(\hat{\theta}) = \left\lbrace \text{diag} \left[ H_T(\hat{\theta})^{-1}B_T(\hat{\theta})H_T(\hat{\theta})^{-1}\right] \right\rbrace ^{1/2}$. Closely related works to ours 
have employed  \eqref{pois-log-lik} for inference ; \cite{paul_etal_2008} and \cite{paul_etal_2011}, among others. However, in such works
\eqref{pois-log-lik} is viewed  as the true log-likelihood of the model
and standard errors are computed  by using  the naive approach  $SE_H(\hat{\theta}) = \left\lbrace \text{diag} \left[ H_T(\hat{\theta})^{-1} \right] \right\rbrace ^{1/2}$ which  underestimates the real source of variation of the parameters when cross-section dependence among counts is present; see \eqref{H_T} which  
depends on the conditional covariance matrix of the process $Y_t$.  The package \texttt{PNAR}  returns  robust standard errors as independence among counts is not assumed for their calculation. 
Similar theory holds for the log-linear model \eqref{log_pnar_pq}; details can be found in the aforementioned works.

\subsection{Influenza data}
\label{subsec: Applied example}

\begin{figure}[H]
	\centering
	\scalebox{0.9}{
		\begin{tabular}{cc}
			\includegraphics[scale = 0.5, trim = 0 0 0 0]{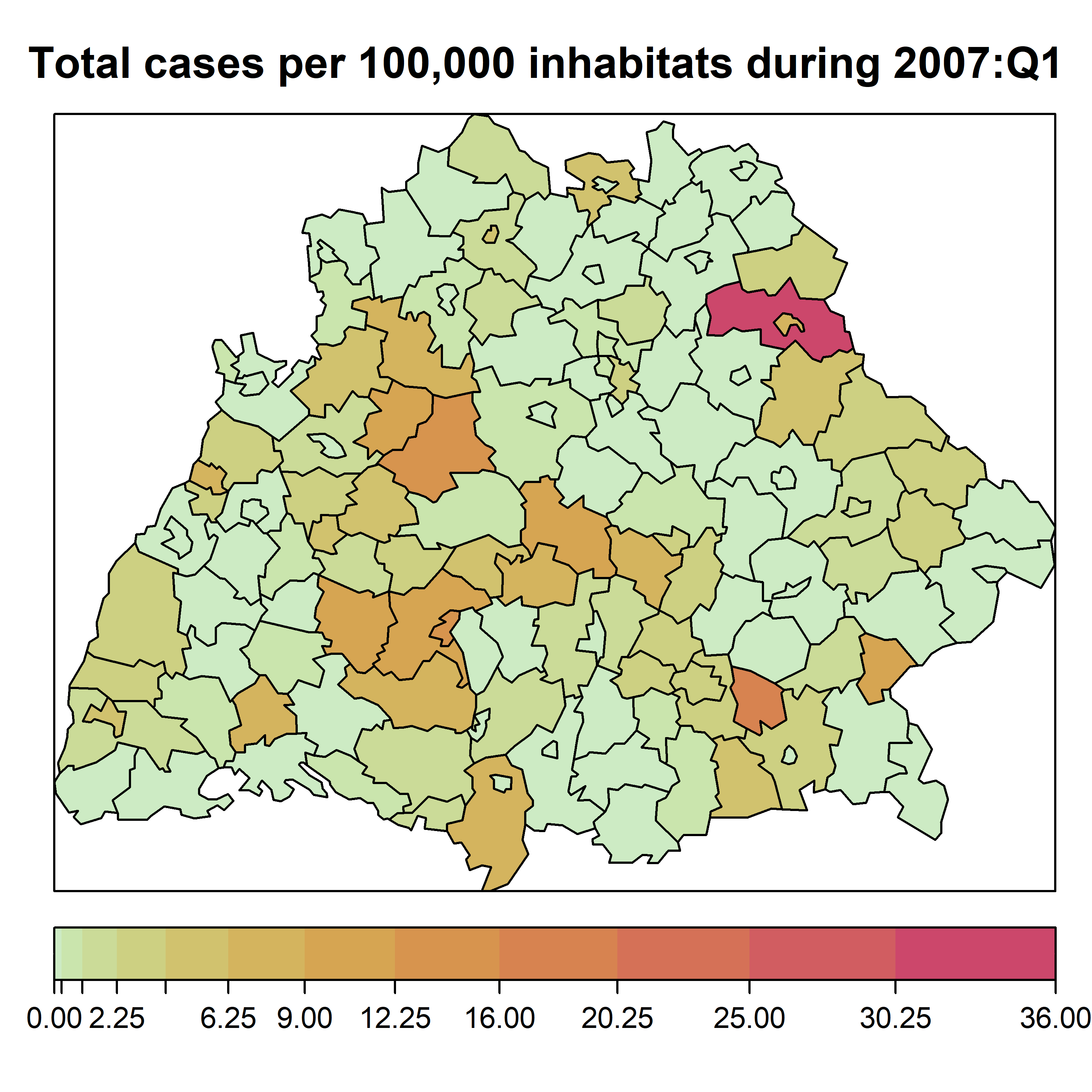} &
			\includegraphics[scale = 0.5, trim = 0 0 0 0]{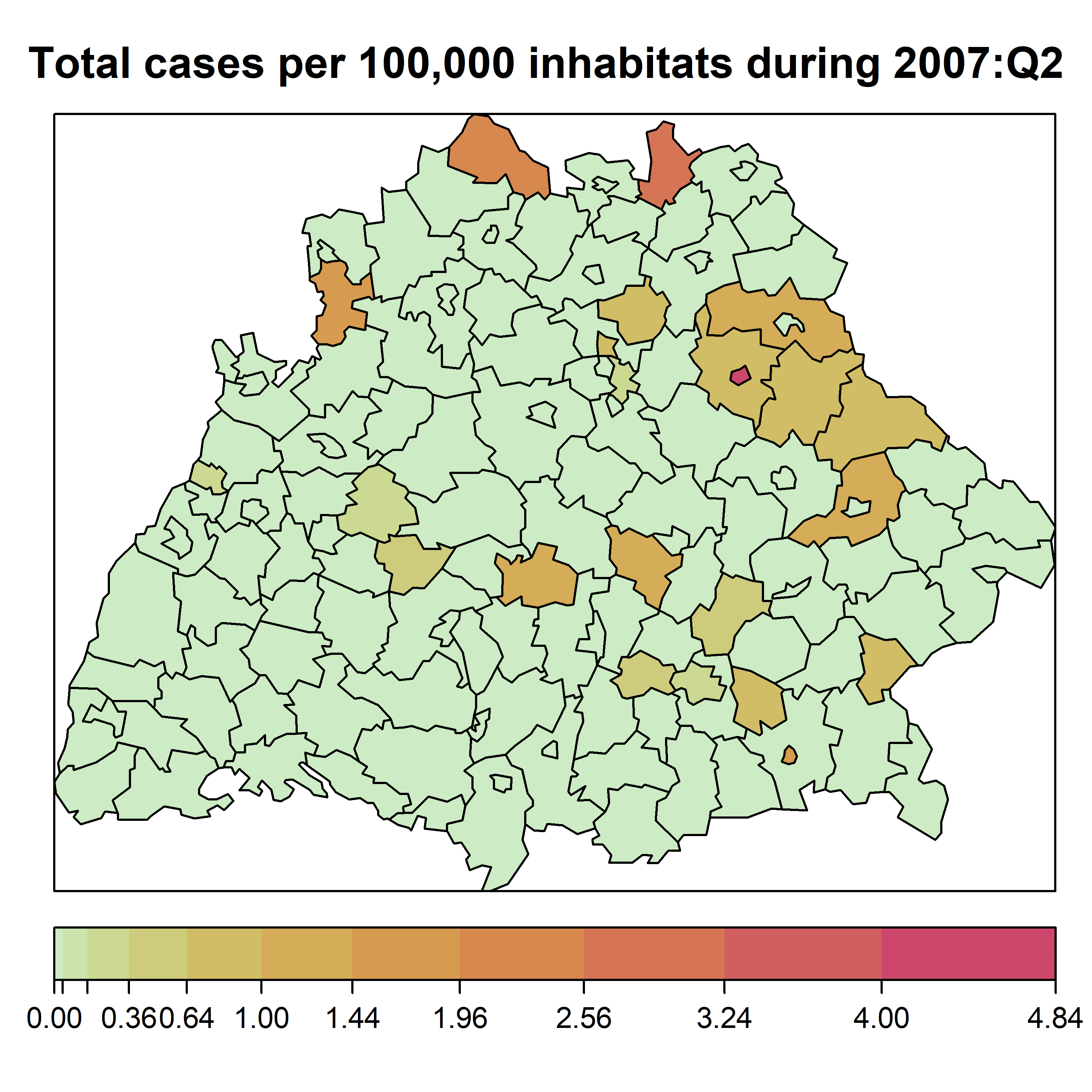} \\
			(a) 1st Quarter of 2007  &  (b) 2nd Quarter of 2007   \\
			\includegraphics[scale = 0.5, trim = 0 0 0 0]{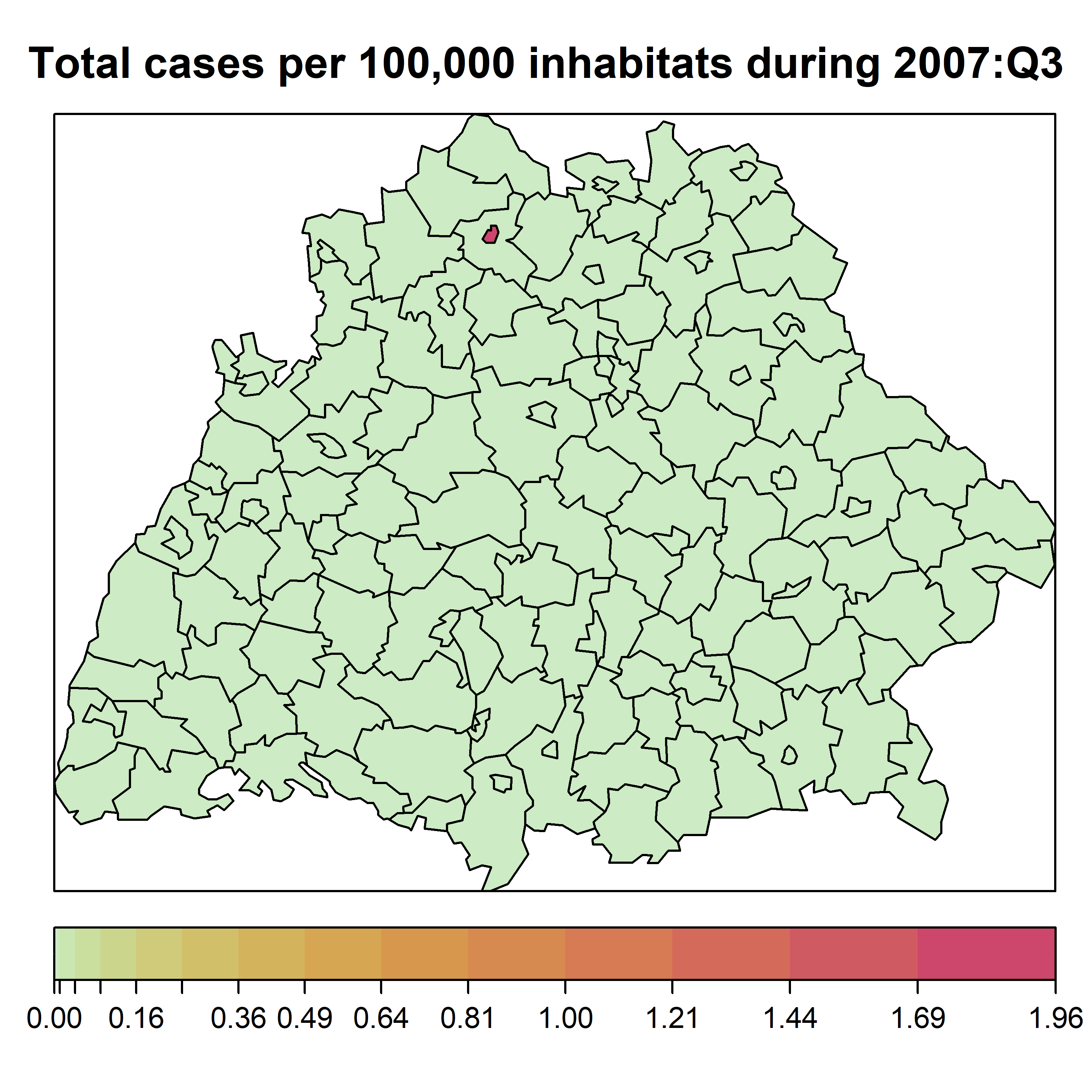} &
			\includegraphics[scale = 0.5, trim = 0 0 0 0]{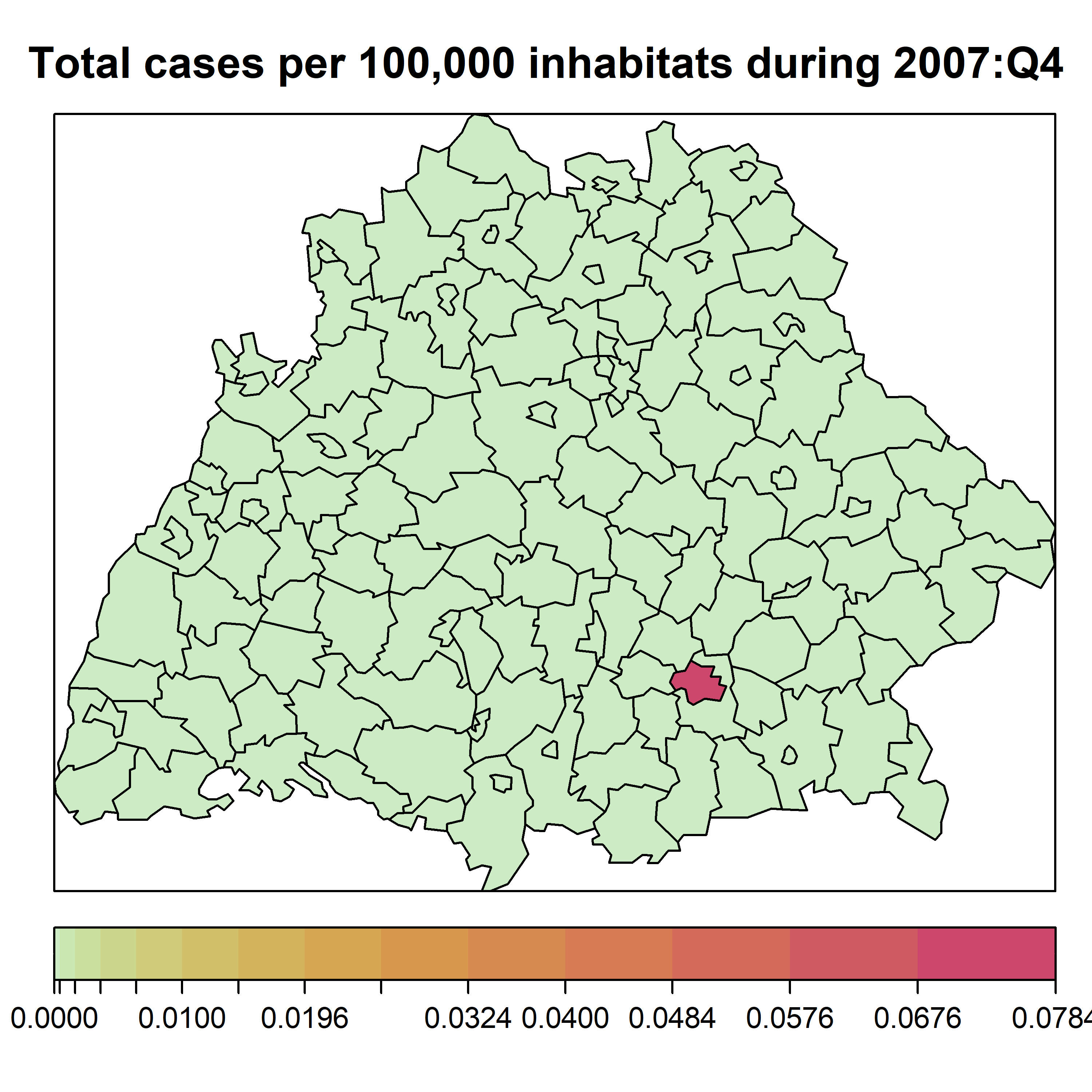} \\
			(a) 3rd Quarter of 2007  &  (b) 4th Quarter of 2007   \\
		\end{tabular}
	}
	\caption{Quarterly flu cases in  two Southern German states Bavaria and Baden-Wuerttemberg for 2007.}
	\label{flu}
\end{figure}

To  illustrate the use of  \texttt{PNAR} we apply the methodology to  the dataset \texttt{fluBYBW} from the \texttt{surveillance} package \citep{Meyer_etal_2017}. This dataset includes information about the weekly number of influenza A \& B cases in the 140 districts of the two Southern German states Bavaria and Baden-Wuerttemberg, for the years 2001 to 2008 (416 time points). The  response variable \break $Y=(Y_{1},\dots,Y_{t},\dots Y_{T})^\prime$ is then a $416 \times 140$ matrix of collective disease counts. Figure~\ref{flu} illustrates  the data in these two regions during 2007.  We model these data by a linear PNAR model as we discuss next.

\begin{verbatim}
	library(PNAR)
	library(surveillance)
	data(fluBYBW)
	flu <- fluBYBW@observed
	A_flu <- fluBYBW@neighbourhood
	pop <- as.matrix(t(fluBYBW@populationFrac)[,1])
\end{verbatim}

\noindent
After loading  \texttt{PNAR}  we  load also \texttt{surveillance}  for obtaining the $140 \times 416$ matrix of collective disease counts \texttt{flu}. The network adjacency matrix \texttt{A\textunderscore flu} of dimension $140 \times 140$ has been obtained by linking two districts if they share (at least) a border.  A covariate vector consisting of  fraction of population in each district is introduced (\texttt{pop}).     
%Model estimation has shown that a PNAR(2) model might be suitable for these data. Indeed, estimation with  $p>2$ did not provide any improvement  (see Table~\ref{tab: 3-4 lags} in the Appendix).
Model estimation for \eqref{pnar_pq} when  $p=1$ and $p=2$ are obtained below  by using the function \texttt{lin\_estimnarpq()} as follows:

\begin{verbatim}
	est1.z <- lin_estimnarpq(y = flu, W = A_flu, p = 1, Z = pop)
	
	est2.z <- lin_estimnarpq(y = flu, W = A_flu, p = 2, Z = pop)
\end{verbatim}

\noindent
Any type of non-negative matrix with zero main diagonal can be used  as a valid adjacency matrix \texttt{W}. For instance consider weighted networks or inverse distance matrices, etc;  see the last \hyperref[sec: Generate data]{section}
for some alternatives when generating data. 
Optimization of \eqref{pois-log-lik} is implemented  under the non-negativity constraint of coefficients  satisfying the  stationary condition. This is a nonlinear constrained optimization problem solved by means of a Sequential Quadratic programming (SQP) gradient-based algorithm \citep{kraft1994algorithm} of package \texttt{nloptr} \citep{nloptr_2022}.
By default, the optimization is constrained in the stationary region; this can be removed by setting the option  \texttt{uncons = TRUE} although this is not suggested because  large sample properties of the estimators have been developed within the stationary region.
The function  \texttt{lin\_estimnarpq()} has three additional features:
\begin{itemize}
	\item \texttt{maxeval}: the maximum number of iterations for the optimization (the default is 100);
	\item \texttt{xtol\textunderscore rel}: relative tolerance for the optimization termination condition (the default is 1e-8);
	\item \texttt{init}: starting value of the optimization (the default is NULL).
\end{itemize}
When \texttt{init = NULL}  starting values are computed internally by ordinary LS  using a lower barrier value of 0.01 because the  regression coefficients cannot assume negative values. However,  users can provide their own initial values through the argument \texttt{init}. 

The function \texttt{lin\textunderscore estimnarpq} returns as output a list consisting of  the estimated coefficients, their associated standard errors, a z-test statistics with $p$-values, the score function evaluated at the optimum, the maximized log-likelihood, and the usual Akaike and Bayesian Information Criteria (AIC, BIC) accompanied by  the Quasi IC (QIC) \citep{pan2001} which takes into account the fact that the log-likelihood \eqref{pois-log-lik} is a quasi log-likelihood; see  Table~\ref{tab:est 1 vs 2} for the results, which are obtained in less than a  second (see also Table~\ref{comp_times}). The score computed at the optimum values is of order 1e-5, on average, indicating  successful convergence of the algorithm.

\begin{table}[ht]
	
	\centering
	\caption{Estimation of linear PNAR model \eqref{pnar_pq} for $p=1,2$ and $Z =$ \texttt{pop}. 
		Standard errors of coefficients are given in parentheses.}
	\scalebox{0.9}{
		\begin{tabular}{rrrrrrrrrr}
			\toprule
			$p$ & $\beta_0$ & $\beta_{1,1}$ & $\beta_{1,2}$ & $\beta_{2,1}$ & $\beta_{2,2}$ & $\delta$ & AIC & BIC & QIC \\ 
			\midrule
			1 & 0.0118 & 0.2862 & - & 0.6302 & - & 2.0027 & -6041.20 & -6025.08 & -5886.16 \\[-0.1cm] 
			& (0.0022) & (0.0204) &  & (0.0345) &  & (0.4475) &  &  &  \\ 
			2 & 0.0081 & 0.2303 & 0.0136 & 0.5459 & 0.1445 & 1.7609 & -7447.48 & -7423.30 & -7240.56 \\[-0.1cm]  
			& (0.0018) & (0.0218) & (0.006) & (0.0379) & (0.0183) & (0.3998) &  &  &  \\ 
			\bottomrule
		\end{tabular}
	}
	\label{tab:est 1 vs 2}
\end{table}
All the estimated coefficients are positive and significantly different from 0. The autoregressive effect $\beta_{2,h}$ shows  higher magnitude with respect to the network effects $\beta_{1,h}$ since  past counts of the same district are, in general, more informative than the neighboring  cases.
Both  network and autoregressive parameters when $p=1$ have a larger  magnitude when compared  to the corresponding coefficients at $p=2$. This can be explained since influenza has an incubation period of only 1-4 days (with an average of 2 days) and a patient is still contagious for no more than 5-7 days after becoming sick\footnote{\href{https://www.cdc.gov/flu/about/disease/spread.htm}{https://www.cdc.gov/flu/about/disease/spread.htm}} so the case counts at first lag are more informative  than the ones at second lag which are still important. 
The population covariate is  significant with positive effect.  Standard errors are computed by using the sandwich estimator.

%\subsection{Lag order selection}
%\label{subsec: Lag order selection}

\begin{figure}[h]
	\centering
	\scalebox{0.9}{
		\includegraphics[scale = 0.7, trim = 0 0 0 0]{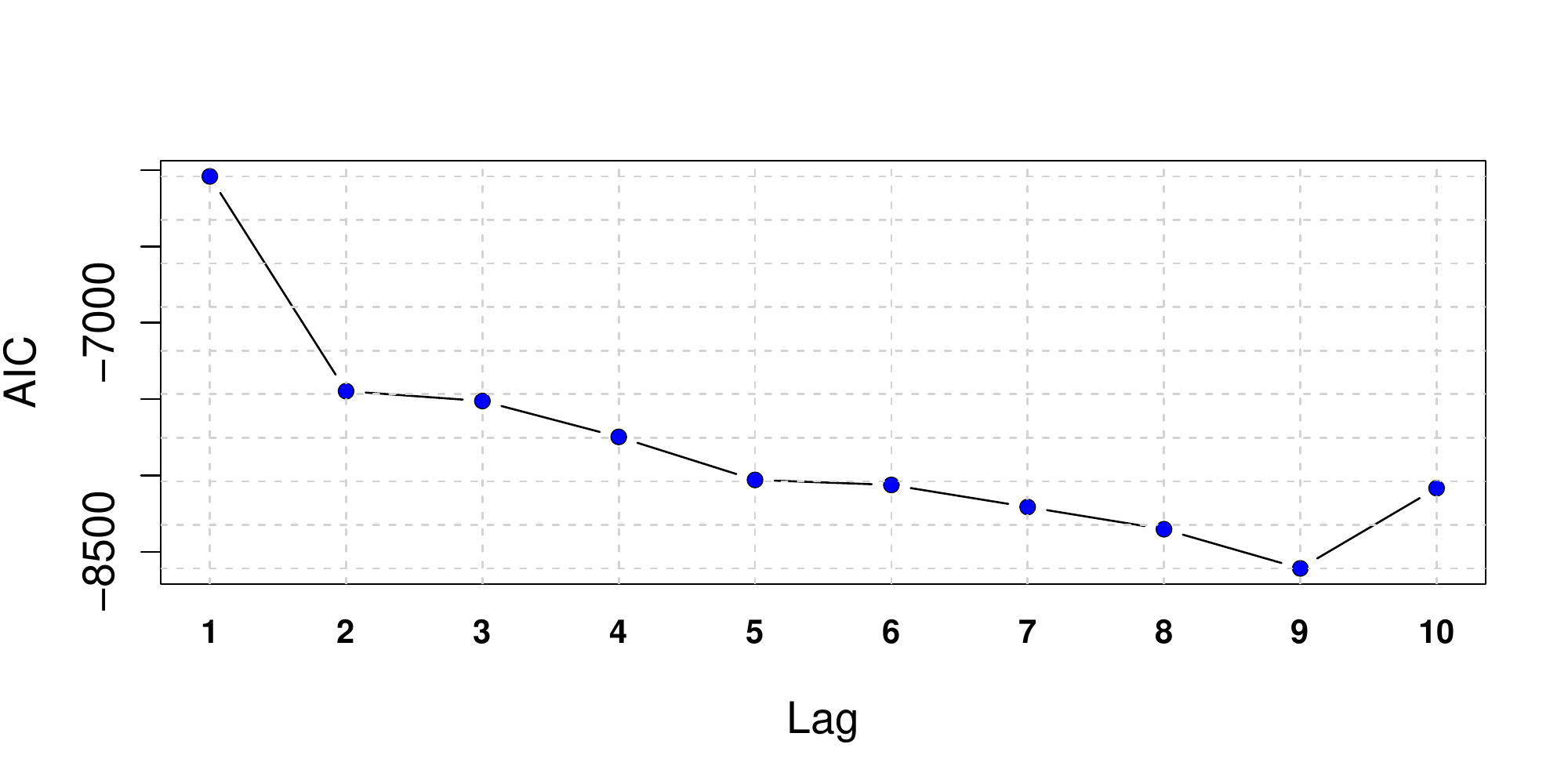}
	}
	\caption{Scatter plot of AIC for PNAR($p$) model versus $p$.}
	\label{ic}
\end{figure}

To select the model order $p$, for the PNAR model \eqref{pnar_pq}, \texttt{PNAR}  includes a function  for estimating model parameters 
for a range of lag values. By  default $p\in \{1,2,\dots,10\}$. This function  returns the scatter plot of any IC (default is QIC) versus the lag order, for example 
\begin{verbatim}
	lin_ic_plot(y = flu, W = A_flu, p = 1:10, Z = pop, ic = "AIC")
\end{verbatim}
Figure~\ref{ic} shows the output of \texttt{lin\_ic\_plot()} for the case of AIC. Plots for the cases of BIC and QIC are similar and not shown. All information criteria  point to the model with $p=9$. However, from the corresponding estimation results reported in the \hyperref[SEC: appendix]{Appendix}, almost all $\beta$ coefficients, which correspond to lags $p \geq 3$, are close to zero and non significant. Therefore, we decide to retain $p=2$, for parsimony. 

We compare  estimation of standard errors after fitting the linear PNAR model to influenza data using \texttt{PNAR} and \texttt{surveillance} packages (covariate $Z$ is excluded). The latter follows the standard error estimation according to the approach described in \cite{paul_etal_2008} and \cite{paul_etal_2011}.
The estimation of model \eqref{pnar_1} with \texttt{PNAR} gives 

\begin{verbatim}
	est1 <- lin_estimnarpq(y = flu, W = A_flu, p = 1)
	summary(est1)
	
	Coefficients: 
	Estimate  Std. Error   z value     Pr(>|z|)    
	beta0  0.02460691 0.002722673  9.037777 1.598906e-19 ***
	beta11 0.28952683 0.020393106 14.197289 9.522500e-46 ***
	beta21 0.63082409 0.034462519 18.304642 7.598940e-75 ***
	--- 
	Signif. codes:  0 '***' 0.001 '**' 0.01 '*' 0.05 '.' 0.1 ' ' 1 
	
\end{verbatim}

\noindent
Using  \texttt{surveillance}, the same model is fitted by the following lines of code:
\begin{verbatim}
	stsObj <- fluBYBW
	ni <- rowSums(neighbourhood(stsObj))
	control <- list( ar = list(f = ~ 1), 
	ne = list(f = ~ 1, weights = neighbourhood(stsObj) == 1, 
	offset = matrix(1/ni, nrow(stsObj), ncol(stsObj), byrow=TRUE)), 
	end = list(f = ~ 1),
	family = "Poisson", keep.terms = TRUE)
	fit1 <- hhh4(stsObj, control)
	coefSE <- coef(fit1, idx2Exp = TRUE, se = TRUE)
	coefSE
	
	Estimate   Std. Error
	exp(ar.1)  0.63082409 0.0066281675
	exp(ne.1)  0.28952683 0.0052551614
	exp(end.1) 0.02460691 0.0007619529
	
\end{verbatim}

\noindent
Comparing the above outputs, 
note that estimated coefficients are identical (subject to rounding errors) but  standard errors, obtained by 
\texttt{surveillance} package,  underestimate their value as it was explained at the end of the inference
\hyperref[subsec:Inference]{section}.
In addition, \texttt{PNAR} is more user-friendly for fitting \eqref{pnar_pq} and it is almost twice as fast. Indeed, 
the average estimation time over 10 calls of the same  function is 0.42 seconds in comparison  to 0.22  by the  \texttt{PNAR} package (see Table~\ref{comp_times}). 

The usefulness of the \texttt{PNAR} package is not limited to epidemiological data. For example, the dataset \texttt{crime}, which is  already built in the package, contains  monthly number of burglaries within the census blocks on the south side of Chicago during 2010-2015 and includes a network matrix, called \texttt{crime\_W}, connecting two blocks sharing a border. The documentation of the
functions \texttt{lin\_estimnarpq()} and \texttt{log\_lin\_estimnarpq()} provides an example of  PNAR models applied to crime data.

\section{Extending linearity}
\label{sec:Linearity test}

In this section, we give  some motivating examples of nonlinear models for time series of networks as  introduced by \cite{armillotta_fokianos_2022_testing}. In addition, we provide testing procedures for testing the linearity assumption.
For ease of presentation, denote by $X_{i,t}=n_i^{-1}\sum_{j=1}^{N}a_{ij}Y_{j,t}$ the average neighbor network mean.
When a drift in the intercept term of \eqref{pnar_pq} is introduced, the following nonlinear Intercept Drift  model, ID-PNAR($p,q$), is obtained
\begin{equation}
	\lambda_{i,t}=\frac{\beta_0}{(1+X_{i,t-d})^{\gamma}}+\sum_{h=1}^{p}\left( \beta_{1h}X_{i,t-h}+\beta_{2h}Y_{i,t-h}\right) + \sum_{l=1}^{q}\delta_lZ_{i,l}\,,
	\label{id_pnar_pq}
\end{equation}
where $\gamma \geq 0$. Model \eqref{id_pnar_pq} shares similar to  a linear model when  the parameter $\gamma$ takes small values   and for $\gamma=0$ reduces  to  \eqref{pnar_pq}. Instead, when $\gamma$ takes values  away from zero, model \eqref{id_pnar_pq} introduces a perturbation, deviating from the linear model, which depends on the network mean at lag $(t-d)$, where $d$ is an additional delay  parameter  such that $d=1,2,\ldots,p$. Model \eqref{id_pnar_pq} is directly applicable  when the baseline effect, $\beta_0$, varies over time as a function of the network. 

Another interesting class of nonlinear models  are regime switching models, i.e. models allowing  for the dynamics of the count process to depend on different regimes (e.g. exponentially expanding pandemic/ dying out pandemic). We give two such examples of models whose specification is based either on a smooth distortion  or an abrupt transition. 
With the same notation as before,  the Smooth Transition PNAR model, ST-PNAR($p,q$) assumes  a smooth transition between  two regimes and it is defined by  
\begin{equation}
	\lambda_{i,t}=\beta_0+\sum_{h=1}^{p}\left( \beta_{1h}X_{i,t-h}+\beta_{2h}Y_{i,t-h} + \alpha_h e^{-\gamma X_{i,t-d}^2}X_{i,t-h}\right) + \sum_{l=1}^{q}\delta_lZ_{i,l} \,,
	\label{stnar_pq}
\end{equation}
where $\gamma \geq 0$ and $\alpha_{h} \geq 0$, for $h=1,\dots, p$. 
This models introduces a smooth regime switching behavior of  the network effect making it possible to vary smoothly  from $\beta_{1,h}$ to $\beta_{1,h} + \alpha_{h}$, as $\gamma$ varies from large to small values. The  additional delay parameter $d$ determines  the time of nonlinear transition that can be chosen.  
When $\alpha_h=0$, for $h=1,\dots,p$ in \eqref{stnar_pq}, the linear PNAR model \eqref{pnar_pq} is recovered. 
In some applications the transition between regimes may be abrupt  (e.g. financial market crashes).
For this reason, we consider the Threshold PNAR model, T-PNAR($p,q$), which is defined by
\begin{equation}
	\lambda_{i,t}=\beta_0+\sum_{h=1}^{p}\left( \beta_{1h}X_{i,t-h}+\beta_{2h}Y_{i,t-h} +\left( \alpha_0 + \alpha_{1h}X_{i,t-h}+\alpha_{2h}Y_{i,t-h}\right) I\left( X_{i,t-d} \leq \gamma \right)  \right) + \sum_{l=1}^{q}\delta_lZ_{i,l}\,,
	\label{tnar_pq}
\end{equation}
where $I(\cdot)$ is the indicator function and $\gamma \geq 0$ is now   threshold parameter. Moreover, $\alpha_{0}, \alpha_{1,h}, \alpha_{2,h} \geq0$, for $h=1,\dots, p$. When $\alpha_{0}=\alpha_{11}=...=\alpha_{2p}=0$, model \eqref{tnar_pq} reduces to \eqref{pnar_pq}. 

For  all nonlinear models, estimation of the unknown parameters is based on  QMLE, following the 
discussion in the inference \hyperref[subsec:Inference]{section}.
Therefore,  analogous conclusions about estimation of regression parameters, their  standard errors and model selection  apply to the case of nonlinear models \eqref{id_pnar_pq}--\eqref{tnar_pq}. Further details can be found in \cite{armillotta_fokianos_2022_testing}. 

\subsection{Standard implementation of testing linearity}
\label{subsec test}

Testing linearity against several specific alternatives  offers guidance about  the type of  nonlinear model to be fitted. Moreover, in certain cases where the linear model is nested within a nonlinear model, some nonlinear parameters may be inconsistently estimated (see the case ST-PNAR and T-PNAR models below) %( see Section~\ref{subsec testnonstandard})
so testing linearity prevents  incorrect estimation.

Consider model \eqref{id_pnar_pq} and  the hypothesis testing problem $H_0: \gamma=0$ vs. $H_1: \gamma>0$ which  is a hypothesis  between the linear PNAR \eqref{pnar_pq} null assumption versus ID-PNAR alternative model \eqref{id_pnar_pq}. Consider the vector of all the parameters of ID-PNAR model \eqref{id_pnar_pq}, $\theta=(\beta_0, \beta_{11},\dots, \beta_{2p}, \delta_1,\dots,\delta_q, \gamma)^\prime$. Define the partition of the parameters $\theta=(\theta^{(1)\prime}, \theta^{(2)\prime})^\prime$, where $\theta^{(1)}=(\beta_0, \beta_{11},\dots, \beta_{2p}, \delta_1,\dots,\delta_q)^\prime$ is the sub-vector of parameters associated 
with the linear part of model.   Let   $\theta^{(2)}$ be the  sub-vector of $\theta$ which corresponds  to  nonlinear parameters; for model \eqref{id_pnar_pq}, $\theta^{(2)}=\gamma$.  Denote further by  $S_{T}(\theta)=( S^{(1)\prime}_{T}(\theta), S_{T}^{(2)\prime}(\theta)) ^\prime$ the  corresponding partition of the quasi score function \eqref{score_poisson}.
We develop a quasi score test statistic based on the quasi log-likelihood \eqref{pois-log-lik}. This is a convenient choice, because  the score test requires  estimation of model under the null hypothesis, i.e. under  the linear model. Then the restricted estimator  is denoted by  $\tilde{\theta}=(\tilde{\beta}_0, \tilde{\beta}_{11},\dots, \tilde{\beta}_{2p}, \tilde{\delta}_1,\dots,\tilde{\delta}_q)^\prime$ and it is usually simpler to compute.  The  quasi score test statistic is given by
\begin{equation}
	LM_{T}= S^{(2)\prime}_{T}( \tilde{\theta}) \Sigma_{T}^{-1}( \tilde{\theta})  S^{(2)}_{T}( \tilde{\theta}) \,,
	\label{score test}
\end{equation}
with $\Sigma_{T}(\tilde{\theta})=\tilde{J} H_T^{-1}(\tilde{\theta}) \tilde{J}^\prime\left( \tilde{J} H_T^{-1}(\tilde{\theta}) B_T(\tilde{\theta}) H_T^{-1}(\tilde{\theta}) \tilde{J}^\prime\right)^{-1} \tilde{J} H_T^{-1}(\tilde{\theta}) \tilde{J}^\prime $, where $\tilde{J}=(O_{m_2\times m_1}, I_{m_2})$, $I_s$ is a $s\times s$ identity matrix and $O_{a\times b}$ is a $a\times b$ matrix of zeros. $\Sigma_{T}(\tilde{\theta})$ is a the estimator for the unknown covariance matrix  
$\Sigma=\mathrm{Var}[S_{T}^{(2)}(\tilde{\theta})]$. It can be proved that the quasi score test \eqref{score test} converges, asymptotically, to a $\chi^2_{m_2}$ distribution, where $m_2$ is the number of nonlinear parameter tested \citep{armillotta_fokianos_2022_testing}. Then, in case of model \eqref{id_pnar_pq}, $m_2=1$ and we can compute the $p$-value of the test statistic \eqref{score test} by 
$p=\mathrm{P}(\chi^2_1\geq M)$, where $M$ is the observed value of the test statistic $LM_T$.  

\subsection{Non-standard implementation of  testing linearity}
\label{subsec testnonstandard}

Suppose now  we wish  to test linearity against \eqref{stnar_pq}. Then, considering the hypotheses 
$ H_0: \alpha_1= \dots = \alpha_p = 0$ versus $H_1: \alpha_h \neq 0$ for some $h=1, \dots, p$,
we note that this testing problem is non-standard  because  it is not possible to estimate the value of  $\gamma$  under  $H_0$. 
Note that the parameter $\gamma$ exists in the partition of the score function \eqref{score_poisson} related to nonlinear parameters, in particular in $\partial\lambda_{i,t}(\theta)/\partial\alpha_h$, and its  associated covariance matrix. Hence, all relevant quantities for computing  \eqref{score test} are  functions of $\gamma$;  that is $S^{(2)}_{T}(\tilde{\theta}, \gamma)$, $\Sigma_{T}(\tilde{\theta}, \gamma)$ and $LM_T(\gamma)$. The model is then subject to non-identifiable parameter $\gamma$ under the null. So if the true model is the linear PNAR model but a ST-PNAR is estimated instead the smoothing parameter $\gamma$ will not be consistently estimated. 
Analogous conclusions hold for testing linearity against the T-PNAR model \eqref{tnar_pq}, where the threshold parameter $\gamma$ is not identifiable under the null. When this issue arises,  the standard theory does not apply and a chi-square type test is not suitable any more; see \cite{davies_1987}, \cite{hansen_1996} and \cite{armillotta_fokianos_2022_testing}, among others. 
It is clear that  the value of the test changes by varying $\gamma\in\Gamma$, where $\Gamma$ is some domain. A summary function of the test, computed under different values of $\gamma$, is then routinely employed in applications; a typical choice is $g_T=\sup_{\gamma\in\Gamma}LM_T(\gamma)$; see  \cite{armillotta_fokianos_2022_testing} who  established the convergence of  $g_T$ to $g=\sup_{\gamma\in\Gamma}LM(\gamma)$, 
where $g$ is a function of a chi-square process, $LM(\gamma)$. The values of the latter asymptotic distribution cannot be tabulated, as they depends on unknown values of $\gamma$.  Hence, we give  methodology for computing $p$-values of  such sup-type test statistic since they cannot be obtained otherwise. 

\paragraph{Davies' bound.} Since the space $\Gamma=[\gamma_L, \gamma_U]$ is usually assumed to be  a closed interval, in practice, we take  $\Gamma_F=(\gamma_L,\gamma_1,\dots,\gamma_l,\gamma_U)$ i.e.  a grid of values for the non-identifiable parameter $\gamma$, and $g_T$ is obtained as the maximum of the tests $LM_T(\gamma)$ computed over $\Gamma_F$. \cite{davies_1987} showed that
\begin{equation}
	\mathrm{P}\left[ \sup_{\gamma\in\Gamma_F} LM_T(\gamma) \geq M\right] \leq \mathrm{P}(\chi^2_{m_2}\geq M)+VM^{\frac{1}{2}(m_2-1)}\frac{\exp(-\frac{M}{2})2^{-\frac{m_2}{2}}}{\Gamma(\frac{m_2}{2})}\,, \label{Davies bound}
\end{equation}
where $M$ is the value of the test statistic $g_T$ computed in the available sample,  $\Gamma(\cdot)$ is the gamma function, and $V$ is the approximated total variation
\begin{equation}
	V=
	|LM_T^{\frac{1}{2}}(\gamma_1)-LM_T^{\frac{1}{2}}(\gamma_L)+|LM_T^{\frac{1}{2}}(\gamma_2)-LM_T^{\frac{1}{2}}(\gamma_1)|+ \dots + |LM_T^{\frac{1}{2}}(\gamma_U)-LM_T^{\frac{1}{2}}(\gamma_l)|. \nonumber
\end{equation}
Equation \eqref{Davies bound} shows how  to approximate the $p$-values of the sup-type test in a straightforward way. Indeed, by adding to the tail probability of a chi-square distribution a correction term, which depends on the total variation of the process, we obtain the desired bound.  This  method is attractive for its simplicity and speed even when the dimension $N$ of the network is large. 
However, the method approximate $p$-values with their bound \eqref{Davies bound} leading to a conservative test. 
In addition,  \eqref{Davies bound} cannot be applied to  the T-PNAR model \eqref{tnar_pq}, because the total variation  requires differentiability of the asymptotic distribution $LM(\gamma)$ under the null hypothesis \cite[p.~36]{davies_1987}, a condition that is not met  for  the case of T-PNAR models. 

\paragraph{Bootstrapping the test statistic.} Based on the previous arguments, we suggest an alternative $p$-value approximation of the test statistic employing stochastic permutations \citep{hansen_1996,armillotta_fokianos_2022_testing}-see Algorithm 1.

\begin{algorithm}
	\caption{Score bootstrap}\label{alg:score}
	\begin{algorithmic}[1]
		\State Obtain the constrained QMLE  of the linear model \eqref{pnar_pq}, say $\tilde{\theta}$ 
		\For{$j=1,\dots,J$}
		\For{$t=1,\dots,T$}
		\State Generate $ \nu_{t,j} \sim N(0,1)$
		\EndFor
		\State Compute $S^{\nu_j}_T(\tilde{\theta}, \gamma)=\sum_{t=1}^{T}s_{t}(\tilde{\theta}, \gamma)\nu_{t,j}$
		\State Compute the test $LM^{\nu_j}_T(\gamma)=S^{\nu_j (2) \prime}_T(\tilde{\theta}, \gamma) \Sigma^{-1}_{T}(\tilde{\theta}, \gamma) S^{\nu_j (2)}_T(\tilde{\theta}, \gamma)$
		\State Optimize $LM^{\nu_j}_T(\gamma)$ for $\gamma$ and take $g^j_T=\sup_{\gamma\in\Gamma}LM^{\nu_j}_T(\gamma)$
		\EndFor
		\State Compute $p^J_T=J^{-1}\sum_{j=1}^{J}I(g^j_T\geq g_T)$
	\end{algorithmic}
\end{algorithm}

An approximation of the $p$-values is obtained from step 10 of Algorithm~\ref{alg:score}, where $g_T$ is the value of the test statistic computed on the available sample.
When  the number of bootstrap replications $J$ is large enough,  $ p^J_T$ provides  a good approximation to  the unknown $p$-values of the test. Then, the null hypothesis $H_0$ is rejected if $p^J_T$ is smaller than a given significance level.

\subsection{Revisiting the  influenza data}
\label{subsec: Applications extensions}

We now apply the testing methodology described in the previous sections to influenza data.
Consider  testing linearity of the  PNAR($2$) model against the nonlinear ID-PNAR($2$)  \eqref{id_pnar_pq}
with $d=1$. Analogous results have been obtained for $d=2$ and therefore are omitted. The quasi score test \eqref{score test}  is computed by

\begin{verbatim}
	id2.z <- score_test_nonlinpq_h0(b = est2.z$coefs[, 1], y = flu, W = A_flu,
	p = 2, d = 1, Z = pop)
	id2.z
	
	Linearity test against non-linear ID-PNAR(p) model
	
	data:  coefficients of the PNAR(p, q)/time series data/order/lag/covariates/
	chi-square-test statistic = 7.2318, df = 1, p-value = 0.007162
	alternative hypothesis: True gamma parameter is greater than 0
	
\end{verbatim}

\noindent
where the first argument requires  estimates  under the null hypothesis $H_0: \gamma = 0$ which have been obtained already. The rest of arguments follow previous syntax. For this testing problem   %(see Sec. \ref{subsec test}) 
the test is asymptotically chi-square distributed with 1 degree of freedom. The output lists  the test statistic value and its corresponding  $p$-value. There is strong indication to  reject  the linear model in favour of  model \eqref{id_pnar_pq}.

Next consider  testing   linearity against the  ST-PNAR($2$) alternative \eqref{stnar_pq}. In this case, the test behaves in a non-standard way  %(see Sec \ref{subsec testnonstandard}) 
so we  will be using the Davies' bound $p$-value (DV) for the sup-type test \eqref{Davies bound} and  the bootstrap $p$-value approximation.  First, the DV  is computed   by calling the following function

\begin{verbatim}
	dv2.z <- score_test_stnarpq_DV(b = est2.z$coefs[, 1], y = flu, W = A_flu,
	p = 2, d = 1, Z = pop)
\end{verbatim}

\noindent
Extreme values  for the range of $\gamma \in \Gamma_F$ are computed internally in such a way that 
$\gamma_U$ and $\gamma_L$ are those  values of $\gamma$ where, on average,  the smoothing function $\exp(-\gamma X_{i,t-d}^2)$ is  equal to 0.1 and 0.9, respectively. In this way, during the optimization procedure, 
the extremes of the function domain are excluded. For more details see the \texttt{PNAR} manual \citep{armillotta_et_al_rpackage_2022}. The user can specify different values for $\gamma_L$ and $\gamma_U$   using the arguments  \texttt{gama\textunderscore L}, \texttt{gama\textunderscore U}  and the number of   grid  values \texttt{len} (the default is 100). Results  suggest  again a deviation from linearity of the model, i.e.

\begin{verbatim}    
	dv2.z
	
	Test for linearity of PNAR(p) versus the non-linear ST-PNAR(p)
	
	data:  coefficients of the PNAR(p, q)/time series data/order/lag/covariates
	/lower gamma/upper gamma/length
	chi-square-test statistic = 35.074, df = 2, p-value = 9.076e-08
	alternative hypothesis: At least one coefficient of the non-linear component 
	is not zero
\end{verbatim}

Next, we apply Algorithm~\ref{alg:score} to compute the  bootstrap $p$-values for the sup-type test statistic. In this case, the observed value of $\sup_{\gamma\in\Gamma_F} LM(\gamma)$ has to be computed.  Initially, perform a global optimization of the $LM_T(\gamma)$ for the ST-PNAR($p$) model, with respect to the nuisance scale parameter $\gamma$ by using Brent's algorithm \citep{brent1972algorithms} in the interval [\texttt{gama\textunderscore L} to \texttt{gama\textunderscore U}],
(see previous discussion for their computation). To ensure global optimality, the optimization is performed on runs at \texttt{len}-1 consecutive equidistant sub-intervals and the global optimum is determined by the maximum over those sub-intervals. The default value  for \texttt{len} is 10. Then using  the function \texttt{global\_optimise\_LM\_stnarpq} with the same arguments we obtain the  optimal $\gamma$ value and the corresponding value of the test statistic: 

\begin{verbatim}
	go1.z2 <- global_optimise_LM_stnarpq(b = est2.z$coefs[, 1], y = flu, W = A_flu,
	p = 2, d = 1, Z = pop)
	go1.z2$gama
	[1] 8.387526
	go1.z2$supLM
	[1] 35.07402
\end{verbatim}

\noindent
This information is used as follows

\begin{verbatim}                  
	boot1.z2 <- score_test_stnarpq_j(supLM = go1.z2$supLM, b = est2.z$coefs[, 1], 
	y = flu, W = A_flu, p = 2, d = 1, Z = pop, 
	J = 499, ncores = 7, seed = 1234)
\end{verbatim}

\noindent
which implements  Algorithm~\ref{alg:score} using $J =499$ bootstrap replicates. The function uses a parallel
processing  option; the user can set the number of cores \texttt{ncores} (the default is no parallel). The seed for random number generation  assures reproducibility of the results.

\begin{verbatim}
	boot1.z2$pJ
	[1] 0.002004008
	boot1.z2$cpJ
	[1] 0.004
\end{verbatim}

\noindent
The above output gives (among other information) $p^J_T$ of step 10 of Algorithm~\ref{alg:score}
and an alternative corrected unbiased estimator for the $p$-value  which is  $cp^J_T=(J+1)^{-1}\left[ \sum_{j=1}^{J}I(g^j_T\geq g_T)+1\right] $. Like in the case of DV $p$-value,  linearity is rejected.

We work analogously  for testing the   PNAR($2$) model versus the T-PNAR($2$) model \eqref{tnar_pq}.
Note that  optimization of $LM_T(\gamma)$, in this case, is based on   \texttt{gama\textunderscore L} and \texttt{gama\textunderscore U} which are obtained (by default) as the mean over $i = 1, \dots,N$ of 20\% and 80\% quantiles of the empirical distribution of the network mean $X_{i,t}$ for $t = 1, \dots,T$. In this way, during the optimization process, the indicator function $I(X_{i,t-d} \leq \gamma)$ avoids values close  to 0 or 1. Alternatively, their value can be supplied by the user. The functions used are analogous to the functions used for the ST-PNAR($2$).

\begin{verbatim}
	tgo1.z2 <- global_optimise_LM_tnarpq(b = est2.z$coefs[, 1], y = flu, W = A_flu,
	p = 2, d = 1, Z = pop)
	tgo1.z2$gama
	[1] 0.1257529
	tgo1.z2$supLM
	[1] 49.06505	
	
	tboot1.z2 <- score_test_tnarpq_j(supLM = tgo1.z2$supLM, b = est2.z$coefs[, 1],
	y = flu, W = A_flu,  p = 2, d = 1, Z = pop,
	J = 499, ncores = 7, seed = 1234)
	tboot1.z2$pJ
	[1] 0.3907816
	tboot1.z2$cpJ
	[1] 0.392
\end{verbatim}

\noindent
The test does not reject the null hypothesis of linearity, in the case of a threshold model. Overall, the analysis shows that the linear PNAR model may not be a suitable  model to fit such epidemic data and nonlinear alternatives should be considered. In particular, evidence of a nonlinear drift in the intercept and of a regime switching mechanism is detected. In addition, it appears that a smooth regime switching mechanism might be more appropriate for the data.

\subsection{Computational speed}

Package \texttt{PNAR} is quite efficient in terms of computational speed, especially for estimation problems. We have employed the \texttt{Rfast} and \texttt{Rfast2} packages \citep{Rfast,Rfast2} wherever possible to ensure computational speed.
We run each function 10 times and compute the average time required to be executed. We use a laptop  computer equipped with  Intel Core i7 processor (3.00GHz) and 16 GB of RAM. Results are given in  Table~\ref{comp_times}.
Estimation of  linear PNAR model  and standard testing (ID-PNAR model) is fast.  
Computations of $p$-values, in the case of non-identifiable parameters,   requires several evaluations of the test statistic on a grid of values for $\gamma$ (Davies' bound) or global optimization of the test statistic (bootstrap). However both tasks are executed in a satisfactory amount of time. The bootstrap approximation Algorithm \ref{alg:score}  runs slower  but it still executed within satisfactory time limits.  Computational  speed, for both Davies' bound and bootstrap $p$-values, can be further increased by reducing the length of the $\gamma$ grid for the former  and the number of bootstrap replications for the latter.  Increasing the number of cores will provide faster bootstrap approximated $p$-values. 

\begin{table}[H]
	\centering
	\caption{Average computation times (in seconds) for the functions called in the text.}
	\begin{tabular}{rrrrr}
		\toprule
		\multicolumn{5}{c}{PNAR Estimation}  \\
		\midrule
		%\cline{2-5}
		$p$ & 1 & 2 & 4 & 9  \\ %3
		\midrule
		No cov. & 0.22  & - & - & - \\
		Cov. & 0.39 & 0.57 & 0.89 & 1.26 \\ % 0.73
		\midrule
		\multicolumn{5}{c}{Tests}  \\
		%\cline{2-5}
		\midrule
		Models & $\chi^2_1$ & $DV$ & Global opt. & Bootstrap \\ \\
		ID-PNAR & 0.43 & - & - & - \\
		ST-PNAR & - & 12.88 & 3.74 & 74.60 \\
		T-PNAR & - & - & 1.30 & 75.20 \\
		\bottomrule
		
	\end{tabular}
	\label{comp_times}
\end{table}

\section{Simulating network count time series}
\label{sec: Generate data}

In this last section we present a further novel implementation of the \texttt{PNAR} package 
which can be used to simulate  network count time series from linear and nonlinear models with multivariate copula Poisson distribution, as we explain next. 

Equation \eqref{pnar_1} does not include information about the joint dependence structure of the PNAR(1) model. 
Following \cite{fok2020} the joint multivariate distribution of the  vector count time series $Y_t$ is defined as $Y_t=N_t(\lambda_t)$ where, $\left\lbrace N_t \right\rbrace $ is a sequence of independent $N$-variate copula-Poisson processes, that is   $N_t(\lambda_t)$ is a sequence of $N$-dimensional IID  marginally Poisson count processes, with intensity 1, counting the number of events in the interval of time $[0,\lambda_{1,t}]\times\dots\times[0,\lambda_{N,t}]$, and whose structure of dependence is modeled through a copula construction $C(\dots; \rho)$ on their associated exponential waiting times random variables. The algorithm is described below for model \eqref{pnar_1}.

Consider a network matrix $A$ and a set of values $(\beta_0,\beta_1, \beta_2)^\prime$ for model \eqref{pnar_1}. Moreover, define a starting mean vector at time $t=0$, say $\lambda_0=(\lambda_{1,0},\dots,\lambda_{N,0})^\prime$.
\begin{enumerate}	
	\item Let $U_{l}=(U_{1,l},\dots,U_{N,l})^\prime$, for $l=1,\dots,K$ a sample from a $N$-dimensional copula \break $C(u_1,\dots, u_N; \rho)$, where $U_{i,l}$ follows a Uniform(0,1) distribution, for $i=1,\dots, N$.
	\item The transformation $E_{i,l}=-\log{U_{i,l}}/\lambda_{i,0}$  follows the exponential distribution  with parameter $\lambda_{i,0}$, for $i=1,\dots, N$.
	\item If $E_{i,1}>1$, then $Y_{i,0}=0$, otherwise 
	$Y_{i,0}=\max\left\lbrace k\in[1,K]:  \sum_{l=1}^{k}X_{i,l}\leq 1\right\rbrace$, by taking $K$ large enough.
	Then, $Y_{i,0}|\lambda_0 \sim Poisson(\lambda_{i,0})$, for $i=1,\dots, N$. So, $Y_{0}=(Y_{1,0},\dots, Y_{N,0})^\prime$ is a set of (conditionally) marginal Poisson processes with mean $\lambda_0$. 
	\item By using the model \eqref{pnar_1}, $\lambda_1$ is obtained.
	\item Return back to step 1 to obtain $Y_1$, and so on.
\end{enumerate}
In  applications, choose  $K$ large , e.g. $K=100$; its value clearly depends, in general, on the magnitude of observed data. Moreover, the copula
$C(\dots; \rho)$ depends on one or more unknown parameters, say $\rho$, which capture the contemporaneous correlation among the variables. The proposed algorithm ensures that all marginal distributions of $Y_{i,t}$ are univariate Poisson, conditionally to the past, as described in \eqref{pnar_1}, while it introduces an arbitrary
dependence among them in a flexible and general way by the copula construction through the parameter $\rho$. An analogous process is employed for generating log-linear and nonlinear count network models by suitable modifications. 

Multivariate Poisson-type distributions have typically complicated form and their  covariance matrix might not be appropriate  \citep{fok2020}; this inspired the adoption of this simulation methodology. Imposing a copula  directly on Poisson marginals can lead to identifiability issues \citep{GenestandNeslehova(2007)}.  For further details see \cite{fok2020}, \cite{armillotta_fokianos_2021} and the recent review in \cite{fokianos_2021}.

\texttt{PNAR} allows to generate multivariate count times  from well-known network models like the Erd\H{o}s-R\'{e}nyi Model \citep{erdos_1959}, with the function \texttt{adja\textunderscore gnp()}, or the Stochastic Block Model (SBM) \citep{wang1987} with the function \texttt{adja()}. Such functions are based on the \texttt{igraph} package \citep{csardi_2006}; see  \cite{armillotta_et_al_rpackage_2022} for details.

%\begin{verbatim}
%	W_SBM <- adja(N = 10, K = 2, alpha = 0.7, directed = TRUE)
%	
%	poisson.MODpq(b = c(0.2,0.2,0.4), W = W_SBM, p = 1, TT = 100, N = 10, 
%	copula = "gaussian", corrtype = "equicorrelation", rho = 0.5)
%	
%\end{verbatim}

\begin{verbatim}
	set.seed(1234) 
	
	W_SBM <- adja(N = 10, K = 2, alpha = 0.7, directed = TRUE)
	
	sim1 <- poisson.MODpq(b = c(0.2,0.2,0.4), W = W_SBM, p = 1, TT = 100, N = 10, 
	copula = "gaussian", corrtype = "equicorrelation", rho = 0.5)
	sim1$y
	
\end{verbatim}

\begin{figure}[h]
	\centering
	\scalebox{0.8}{
		\includegraphics[scale = 0.8, trim = 0 0 0 0]{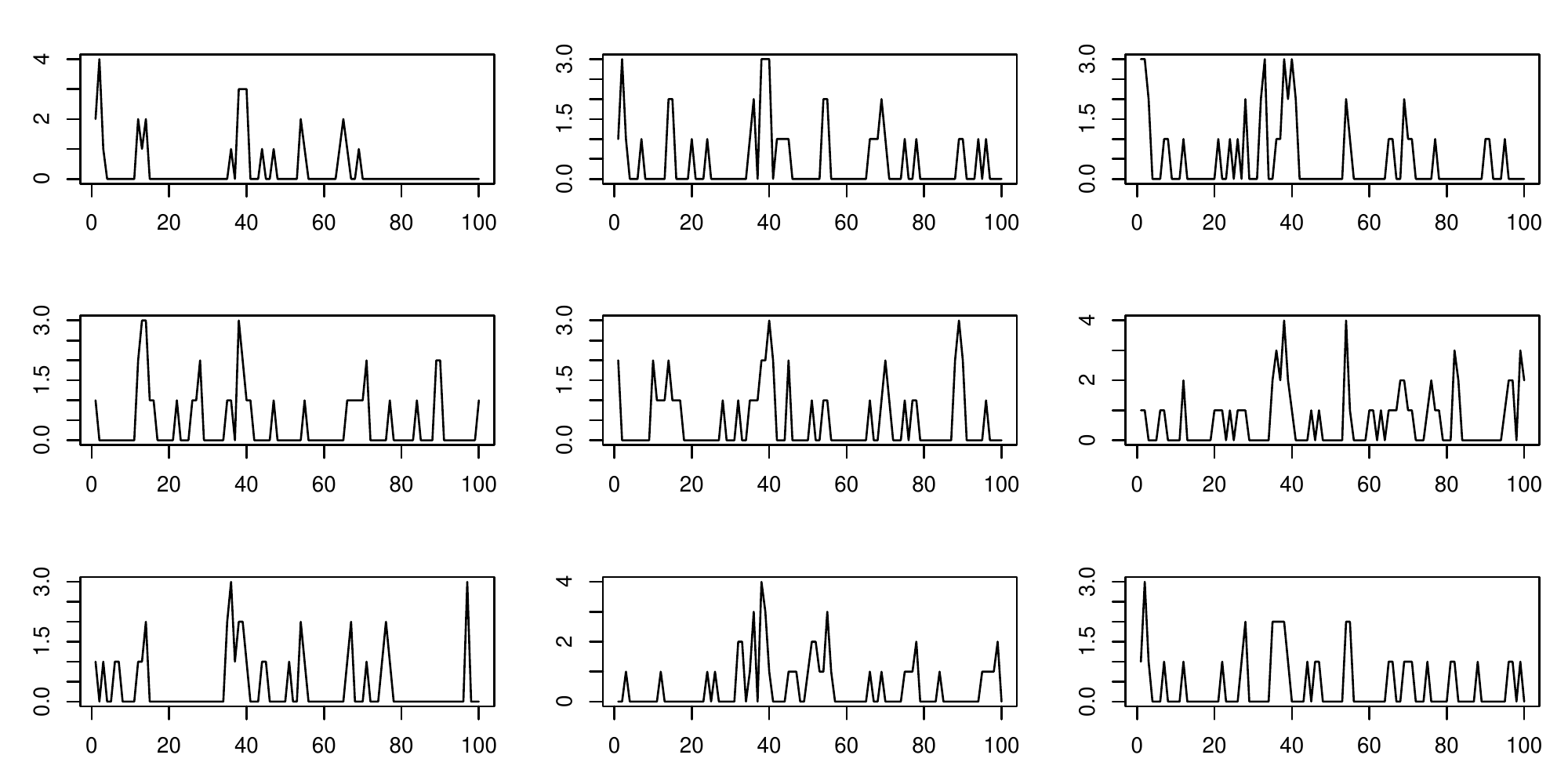}
	}
	\caption{Simulated count time series from the linear PNAR(1) model.}
	\label{simulation}
\end{figure}

\noindent
The first function randomly generates an adjacency matrix from the directed SBM model with 10 nodes and 2 groups. The second function generates a $100 \times 10$   time series matrix object of network counts from the linear PNAR(1) model \eqref{pnar_1} where the joint dependence in the data generating process is modeled by a Gaussian copula with $\rho=0.5$. The "equicorrelation" option generates a correlation matrix for the Gaussian copula where all the off-diagonal entries equal $\rho$. Another type of correlation matrix which  can be used is the "toeplitz" option that returns  a correlation matrix whose generic off-diagonal $(i, j)$-element is $\rho^{|i-j|}$. Moreover, other copula functions can also be chosen as the $t$ or the Clayton copula. Some of the 10 simulated time series are plotted in Figure~\ref{simulation}, for illustration. Analogous functions are provided for 
generating synthetic data from log-linear or nonlinear Poisson network model (Table~\ref{functions2} in the Appendix).

\section{Conclusion}
\label{sec. Discussion}

There exists R software for fitting Network Autoregressive models (\texttt{GNAR}, \texttt{surveillance}). However, no published package includes functions for  inference with nonlinear Network Autoregressive models. 
\texttt{PNAR} fills this  gap by providing users tools for  efficient estimation of (log-)linear models with proper robust 
standard errors, test statics and computational algorithms  for  testing model linearity 
and new simulation methodology for generating  (log-)linear and nonlinear network count time series models. 
We showed that all these tasks are executed  with a minimal computational effort.

There are a number of possible developments for \texttt{PNAR}. One possibility is to include several other nonlinear models and develop related linearity tests. In addition, developing a negative binomial quasi-likelihood estimation method offers more flexibility to model fitting.  Alternative ways to compute $p$-values, for example employing different bootstrap approximation, may also be considered.  Further simulation methods for generating network count time series are easily accommodated by suitable modification of the copula Poisson algorithm.  All these extensions provide users with new set of tools for inference in the broad framework of multivariate discrete-valued time series models.

\section{Appendix}
\label{SEC: appendix}

%\begin{table}[H]
%	\centering
%	\caption{Estimation of linear PNAR model \eqref{pnar_pq} for $p=3,4$ and $Z = $ \texttt{pop}. 
	%		Standard errors of coefficients are given in parentheses.}
%	\scalebox{0.8}{
	%		\begin{tabular}{ccccccccccc}
		%			\toprule
		%			$p$ & $\beta_0$ & $\beta_{1,1}$ & $\beta_{1,2}$ & $\beta_{1,3}$ & $\beta_{1,4}$ & $\beta_{2,1}$ & $\beta_{2,2}$ & $\beta_{2,3}$ & $\beta_{2,4}$ & $\delta$ \\ 
		%			\midrule
		%			3 & 0.0084 & 0.2304 & 0.0103 & 0.0000 & - & 0.5453 & 0.1363 & 0.0134 & -  & 1.6525 \\
		%			& (0.0019) & (0.0218) & (0.0055) & (0.0033) &  & (0.0382) & (0.0179) & (0.0052) &  & (0.3958) \\ 
		%			4 & 0.0083 & 0.2303 & 0.0109 & 0.0000 & 0.0000 & 0.5453 & 0.1364 & 0.0136 & 0.0000 & 1.5711 \\ 
		%			& (0.0019) & (0.0217) & (0.0056) & (0.0034) & (0.0046) & (0.0382) & (0.0177) & (0.0052) & (0.002) & (0.3816) \\ 
		%			\bottomrule
		%		\end{tabular}
	%	}
%	\label{tab: 3-4 lags}
%\end{table}

Output from estimation of PNAR model \eqref{pnar_pq} with lag $p=9$ and population covariate:
\begin{verbatim}
	est9.z <- lin_estimnarpq(y = flu, W = A_flu, p = 9, Z = pop)
	summary(est9.z)
	
	Coefficients: 
	Estimate   Std. Error      z value     Pr(>|z|)    
	beta0  8.046948e-03 0.0018543147 4.339581e+00 1.427548e-05 ***
	beta11 2.291637e-01 0.0222504112 1.029930e+01 7.097436e-25 ***
	beta12 1.219436e-02 0.0058931749 2.069234e+00 3.852414e-02 *  
	beta13 2.880415e-08 0.0034601665 8.324498e-06 9.999934e-01    
	beta14 6.840207e-08 0.0047642557 1.435735e-05 9.999885e-01    
	beta15 4.994822e-08 0.0022529259 2.217038e-05 9.999823e-01    
	beta16 6.335964e-07 0.0023712114 2.672037e-04 9.997868e-01    
	beta17 1.096226e-06 0.0022836178 4.800390e-04 9.996170e-01    
	beta18 1.094862e-07 0.0018014328 6.077731e-05 9.999515e-01    
	beta19 1.230050e-07 0.0017010627 7.231070e-05 9.999423e-01    
	beta21 5.462136e-01 0.0389427608 1.402606e+01 1.079843e-44 ***
	beta22 1.361396e-01 0.0179869627 7.568794e+00 3.767043e-14 ***
	beta23 1.404481e-02 0.0053121667 2.643894e+00 8.195825e-03 ** 
	beta24 1.764756e-08 0.0023065543 7.651051e-06 9.999939e-01    
	beta25 1.874096e-06 0.0016152031 1.160285e-03 9.990742e-01    
	beta26 2.036961e-05 0.0023094311 8.820187e-03 9.929626e-01    
	beta27 1.492721e-06 0.0019472654 7.665731e-04 9.993884e-01    
	beta28 6.414357e-08 0.0018167793 3.530620e-05 9.999718e-01    
	beta29 8.818047e-07 0.0009314276 9.467238e-04 9.992446e-01    
	delta1 1.412017e+00 0.3536615653 3.992566e+00 6.536216e-05 ***
	--- 
	Signif. codes:  0 '***' 0.001 '**' 0.01 '*' 0.05 '.' 0.1 ' ' 1 
	
	Log-likelihood value:  4324.057 
	AIC: -8608.114 BIC: -8527.5 QIC: -8362.176
\end{verbatim}

\begin{table}[H]
	\caption{A brief overview of the main functions in \texttt{PNAR}. \label{functions} }
	\centering
	\scalebox{0.9}{
		\begin{tabular}{rr} 
			\toprule
			\textbf{Function}              &  \textbf{Description}           \\     \midrule
			\texttt{lin\textunderscore estimnarpq()}      &  Fitting the linear PNAR($p,q$) model.      \\  
			\texttt{log\textunderscore lin\textunderscore estimnarpq()}  &  Fitting the log-linear PNAR($p,q$) model. \\ \texttt{score\textunderscore test\textunderscore nonlinpq\textunderscore h0()} & Score test of linear PNAR versus ID-PNAR model. \\
			\texttt{score\textunderscore test\textunderscore stnarpq\textunderscore DV()} &  Score test of linear PNAR versus ST-PNAR model by \eqref{Davies bound}. \\
			\texttt{global\textunderscore optimise\textunderscore LM\textunderscore stnarpq()} & Maximize test statistic of ST-PNAR for nuisance parameters. \\
			\texttt{score\textunderscore test\textunderscore stnarpq\textunderscore j()} & Bootstrap score test of linear PNAR versus ST-PNAR model.\\
			\texttt{global\textunderscore optimise\textunderscore LM\textunderscore tnarpq()} & Maximize test statistic of T-PNAR for nuisance parameters. \\
			\texttt{score\textunderscore test\textunderscore tnarpq\textunderscore j()} & Bootstrap score test of linear PNAR versus T-PNAR model. \\  \bottomrule 
		\end{tabular}
	}
\end{table}

\begin{table}[H]
	\caption{A brief overview of functions simulating network count time series  models in \texttt{PNAR}. \label{functions2} }
	\centering
	\scalebox{0.9}{
		\begin{tabular}{rr} 
			\toprule
			\textbf{Function}                &  \textbf{Description}           \\     \midrule
			\texttt{poisson.MODpq()}  &  Generation from  linear PNAR($p$) model with covariates.      \\  
			\texttt{poisson.MODpq.log()}  &  Generation from  log-linear PNAR($p$) model with covariates.      \\  
			\texttt{poisson.MODpq.nonlin()}  &  Generation from  Intercept Drift PNAR($p$) model with covariates.      \\  
			\texttt{poisson.MODpq.stnar()}  &  Generation from   Smooth Transition PNAR($p$) model with covariates.  \\
			\texttt{poisson.MODpq.tnar()} & Generation from  Threshold PNAR($p$) model with covariates. \\  \bottomrule
		\end{tabular}
	}
\end{table}

\section{Acknowledgments}
Part of this work was done while M. Armillotta was with the Department of Mathematics \& Statistics, University of Cyprus.
We cordially thank R. Hyndman and two anonymous reviewers for several constructive comments that improved an earlier version 
of the manuscript. In addition, we thank M. Papadakis for his help with the S3 methods (print() and summary() functions).
M. Armillotta acknowledges financial support from the EU Horizon Europe programme under the Marie Skłodowska-Curie grant agreement No. 101108797.

\bibliographystyle{apalike}
\bibliography{armillotta}

\end{document}